\documentclass[prd,12pt,onecolumn,tightenlines,superscriptaddress,numerical,notitlepage,showpacs,amsmath,amssymb,amsfonts,aps,longbibliography,floatfix
]{revtex4-1}
\usepackage{graphicx}
\usepackage{subcaption}
\usepackage[colorlinks=True,linkcolor=red,citecolor=blue,urlcolor=blue]{hyperref}
\usepackage{bm}
\usepackage{bbm}
\usepackage{array}
\usepackage{braket}
\usepackage{cancel}
\usepackage{xcolor}
\usepackage{slashed}
\usepackage{comment}
\usepackage{combelow}
\usepackage{chngcntr}
\usepackage{nicefrac}
\usepackage{bookmark}
\usepackage[T1, T2A]{fontenc}
\usepackage[english]{babel}
\usepackage[normalem]{ulem}

\newcommand{\ds}{\displaystyle}

\begin{document}
\selectlanguage{english} 

\title{Net-baryon number distributions and the QCD phase diagram}

\author{
2026 \firstname{R.~N.}~\surname{Rogalyov}}
\email[E-mail: ]{rnr@ihep.ru}
\affiliation{NRC ”Kurchatov Institute” - IHEP, 142281 Protvino, Moscow region, Russia}
%
\author{\firstname{N.~V.}~\surname{Gerasimeniuk}}
\affiliation{Pacific Quantum Center, Far Eastern Federal University, Vladivostok 690922, Russia}

\begin{abstract}
\noindent \textbf{Abstract} --- Net-proton  multiplicity distributions obtained by the STAR
collaboration are analyzed in the grand canonical approach. A method of finding the baryon chemical potential $\mu_B$ immediately from such distributions is applied for the first time. The obtained values of $\mu_B$ are consistent both with those determined earlier using the statistical thermal model and with the results of the fit
based on the Hadron Resonance Gas (HRG) model. We propose a criterion of thermalization of fireballs produced in heavy-nuclei collisions.
\end{abstract}

\maketitle
%


\section{Introduction}

In recent decades, significant efforts of the high-energy physics community have been directed towards searching for the chiral phase transition line and the critical end point in the plane of temperature $T$ and $\mu_B$ \cite{Borsanyi:2025ttb}. In this connection, determination of thermodynamical parameters describing the state of fireballs produced in collisions of heavy nuclei is an important task in the hadron and nuclear physics.

It is generally assumed that the state of 
a ﬁreball produced in a high-energy collision
of heavy nuclei is described by 
the grand canonical partition function. 
Thus it is considered that the strong-interacting
matter at the final stage of the fireball 
evolution is thermalized so that thermodynamical parameters such as $T$ and $\mu_B$ make sense.
In this study, we apply the method for determination of $\theta = \mu_B/T$ at the chemical freezout point suggested 
in~\cite{Nakamura:2013ska} 
to the data on the net-proton number distributions obtained in the Beam Energy Scan at RHIC~\cite{STAR:2017sal}.
In the framework of this approach we formulate criterion of fireball thermalization
in terms of the net-proton multiplicity 
distributions.

We also use fit function based on the 
HRG model to find $\theta$ 
and discuss the parameter of this function
obtained in lattice QCD.

\section{Determination of $\mu_B/T$ based on $CP$-parity conservation}

Strong-interacting matter in thermodynamical equilibrium is characterized by the grand canonical partition function satisfying the fugacity expansion
\[ 
Z_{CG}(\theta)=\sum_{B=-\infty}^{+\infty}
Z_C(B)\exp(\theta B)\;,
\] 
where $Z_C(B)$ is the canonical partition function and $B$ is the baryon number. This being so, 
\[ 
 \varphi_B = {Z_C(B) e^{\theta B} \over Z_{GC}(\theta)}
\] 
provides probability that the baryon number of the system equals $B$. It should be noticed that 
$CP$-transformation changes the sign 
of $B$ and $Z_C(B)=Z_C(-B)$ because 
one can safely neglect $CP$-violation 
in fireball evolution.
Since $Z_C(B)$ is independent of $\theta$, 
we arrive at 
\[  
 {\varphi_B\over \varphi_{-B}} = e^{2\theta B} \quad \implies
 \quad \theta={1\over 2B} \ln \left({\varphi_B\over \varphi_{-B}}\right)
\]  
This formula was suggested in~\cite{Nakamura:2013ska} for extracting 
the values of $\theta$ from experimental data. 
However, the procedure of finding $\theta$ in~\cite{Nakamura:2013ska}
relies on the values of $\varphi_B$
extracted from the data 
on several cumulants with the use of the
probability mass function based on the HRG  model.

Assuming that $\mu_B\approx \mu_P$,
where $\mu_P$ is the chemical potential
associated with the net-proton number $P$,
we find $\mu_B$ from the analysis of 
the net-proton number distribution
rather than from its cumulants.

Here we use the respective data of the STAR  collaboration~\cite{STAR:2021iop}  
so that $\ds \varphi_P={N_P\over N_{tot}}$, 
where $N_P$ is the number of events
with net-proton number $P$ in the specified 
acceptance: rapidity bin is $|y|<0.5$ and 
the transverse momentum varies over the range 
$0.4\div 2.0$~GeV/c. The net-proton  number 
distributions under consideration are shown in Fig.3 of~\cite{STAR:2017sal}.
This being so, the normalized number of events
on the tails of the distribution does not give 
a correct information on the probabilities 
$\varphi_P$ if the expected value of 
$\varphi_P$ falls below some 
value of the order $\ds \sim 1/N_{tot}$,
a robust estimate of the error in $\ln \varphi_P$ 
may exceed $\ln (\varphi_P N_{tot})$. 
For this reason, we discard these data points,  
actually these are $1\div 2$ data points 
at the tails of the empirical distribution function.

\begin{figure}[!ht]
\includegraphics[width=0.6\linewidth]{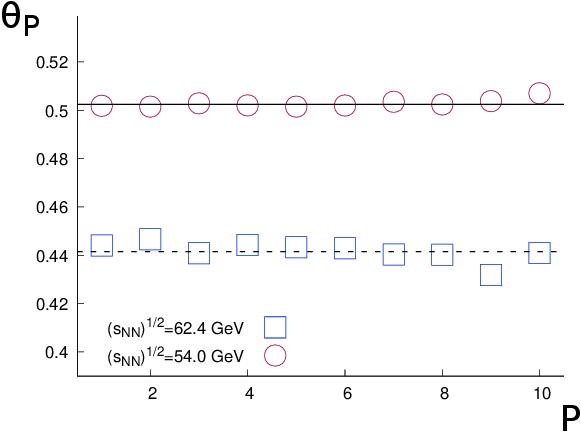}
\caption{Values of $\theta_P$ 
  computed by formula (\ref{eq:1})
  for $\sqrt{s_{NN}}=54.4$ and 62.4~GeV. The results 
  for $P$ on the tails of the net-proton number empirical distribution are not shown.}
\label{fig:ImgA}
\end{figure}

The remaining data can be used to check 
whether the fireballs are in thermodynamical equilibrium.
We suggest a criterion of thermalization: the distribution $\varphi_P$ in the net-proton number $P$ should be such that 
\begin{equation}\label{eq:1}
\theta_P = {1\over 2P} \ln \left( \varphi_P \over \varphi_{-P} \right) 
\end{equation} 
is independent of $P$; in this case 
the estimate of the dimensionless baryon chemical 
potential in the grand canonical approach based 
on $CP$-conservation in fireball evolution is 
given by the formula
\begin{equation} \label{2}
\theta_{CP}\equiv {\langle \theta \rangle} =
{1\over N} \sum_{P=1}^N \theta_P\;,
\end{equation}
where $N$ is the number of data points 
with negative values of $P$ taken into account 
in our analysis.

The quantitative formulation of the thermalization 
criterion is as follows: {\it (i)} the deviations of $\theta_P$ from $\langle \theta \rangle$
must be significantly less than one and {\it (ii)}
no regular dependence of $\theta_P$ on $P$ is observed. 
Thus we consider that the fireball to be 
in thermodynamic equilibrium if, first,
$\delta \theta\ll 1$, where $\delta \theta$ 
is the square root of the variance of $\theta_P$ 
and, second, the coefficient of determination in 
a regression model based on a low-degree 
polynomial in the independent variable 
$P$ vanishes. In Fig.\ref{fig:ImgA}, 
the values of $\theta_P$ are shown as the function 
of $P$ for $\sqrt{s_{NN}}=54.4$ and 62.4~GeV, 
the best regression model is provided by a constant, 
a linear function does not yield a statistically significant decrease of the residual sum of squares 
giving evidence for thermodynamic equilibrium 
of the system under study.

In this case, deviations of $\theta_P$ from 
$\langle \theta \rangle$ should be considered 
as random fluctuations, and $\delta \theta$ 
provides the error in $\theta_{CP}$. 
The method of determination of $\theta_{CP}$ that
we use works only if there is significant 
statistics providing sufficient number 
of data points at $P<0$. For this reason 
$\theta_{CP}$ at $\sqrt{s_{NN}}< 19$~GeV cannot be evaluated.
Our results for $\mu_{CP}=\theta_{CP} T$ are 
presented in Table~\ref{tab:1}, the respective 
values of $T$ are taken from~\cite{STAR:2017sal}. 

\begin{table}[h]
\caption{Results for the baryon chemical potential 
$\mu_{CP}\equiv \theta_{CP} T$ are compared with 
$\mu_{STAR}$. The difference between the values of $\mu_{CP}$ in 3rd and 5th columns is explained by different values of $T$ in the GCER and GCEY approaches.}
 \begin{tabular}{|c|c|c|c|c|} \hline
 $\sqrt{s_{NN}}$, GeV  & $\mu_{STAR}$,~MeV  &  $\mu_{CP}$,~MeV & $\mu_{STAR}$,~MeV  &  $\mu_{CP}$,~MeV   \\   
 & GCER & GCER & GCEY & GCEY      \\ \hline
19.6   & 187.9(8.6) & 175.8(7.8)  & 195.6(9.7) & 179.9(6.3)  \\ 
27.0   & 144.4(7.2) & 140.4(4.6)  & 151.9(9.3) & 144.7(2.8) \\ 
39.0   & 103.2(7.4) & 103.6(3.6)  & 104.7(11.2) & 105.9(2.3) \\ 
62.4   &  69.8(5.6) &  69.9(2.2)  & 69.2(11.4) & 71.6(1.6) \\ 
200.0  &  28.4(5.8) &  27.3(1.0)  & 27.0(11.4) & 27.9(0.8) \\ \hline
 \end{tabular}
 \label{tab:1}
\end{table}

They are compared with the 
values of ${\mu_B}$ evaluated in  \cite{STAR:2017sal}
on the basis of the statistical thermal model~\cite{Wheaton:2009chw} in the framework of the 
Grand Canonical Ensemble Yields (GCEY) 
and Ratios (GCER) approaches. 
The value of $\mu_B$ obtained from statistical 
thermal model analyses \cite{Wheaton:2009chw,STAR:2017sal} 
is designated by $\mu_{STAR}$. 
The excellent agreement between the results 
presented in Table~\ref{tab:1} and obtained by 
fundamentally different methods 
demonstrates reliability of the method for 
determining $\mu_B$ based on formulas 
(\ref{eq:1}) and (\ref{2}). It should be noticed that we did not take into account the correction for the net-baryon number conservation and other listed in \cite{Gazdzicki:2026ubm}.

\section{Comparison with the hadron resonance gas model}
 
\begin{figure}
\includegraphics[width=0.6\linewidth]{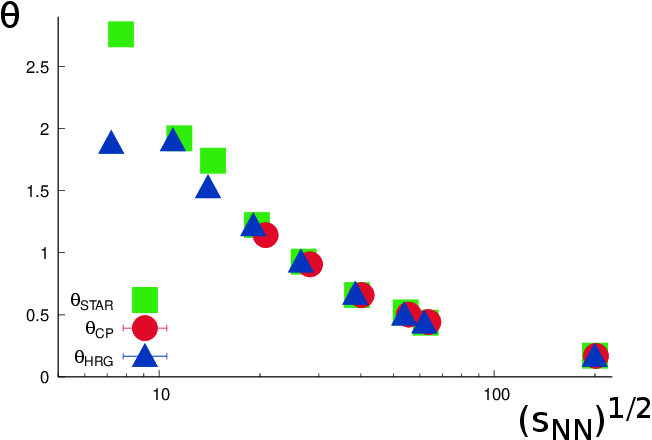}
\caption{The values of dimensionless
 baryon chemical potential, $\theta_{HRG}$, 
 obtained from the fit of the HRG-model formula to the STAR data, are compared with $\theta_{CP}$ and $\theta_{STAR}$.}
\label{fig:ImgB}
\end{figure}

The method of extracting $\mu_B$ from 
experimental data described in the previous 
Section is model-independent: it relies 
only on general assumptions.
Therewith, the net-baryon number distributions 
are predicted both in lattice QCD and in various models. The most popular 
is the HRG model, its basics can be found in the introduction 
to~\cite{Wheaton:2009chw}. It was shown~\cite{Braun-Munzinger:2003pwq, Braun-Munzinger:2011shf} that the 
HRG net-baryon probability distribution can be expressed in terms of average values of the number of baryons $b$ 
and antibaryons $\bar b$ (so that $B=b-\bar b$) as follows:
\begin{equation}\label{eq:3}
    \varphi_B = \left( {\langle b \rangle\over \langle \bar b \rangle} \right)^{B/2} I_B\left(2\sqrt{\langle b \rangle\langle \bar b \rangle}\right) \exp\left[ -( \langle b \rangle + \langle \bar b \rangle )\right] 
\end{equation}
where $I_n(z)$ is the Macdonald function.
In the HRG model, the condition for the coincidence 
of the baryon and proton chemical potentials, 
\begin{equation}\label{eq:4}
 {\langle p \rangle\over \langle \bar p \rangle} =  {\langle b \rangle\over \langle \bar b \rangle} = \exp(2\theta) \,,
\end{equation}
is valid if the isospin chemical potential vanishes, which is a good approximation at 
$\sqrt{s_{NN}}>10$~GeV~\cite{Kitazawa:2012at}.
Formula (\ref{eq:3}) represents the well known Skellam distribution, it can also be written in terms of variables $a=2\sqrt{\langle b \rangle \langle \bar b \rangle}$ and $ \theta$.
Thus the distribution in net-proton number $P$ is given by
\begin{equation}\label{eq:5}
    \varphi_P= I_P(A) \exp\left( P \theta  - A \cosh (\theta)\right)\;,
\end{equation}
where 
\[ 
A=2\sqrt{\langle p \rangle \langle \bar p \rangle}\;,
\]  
$p$ and $\bar p$ are the numbers of protons 
and antiprotons, respectively.
Formula (\ref{eq:5}) provides a fit function 
for the data under consideration. The values of 
$\theta$ determined from the fit of the function 
(\ref{eq:5}) to the data are designated 
by $\theta_{HRG}$ and shown in Table~\ref{tab:2} 
and in Fig.\ref{fig:ImgB}.

\begin{table}
\caption{The values of $\theta$ presented in  \cite{STAR:2021iop} ($\theta^a_{STAR}$) and 
\cite{Nakamura:2013ska} ($\theta^{\rm{old}}_{CP}$) are compared 
with our values obtained by formulas (\ref{eq:1}--\ref{2})
($\theta_{CP}$) as well as by the HRG fit function 
($\theta_{HRG}$) at various energies of collision.  Quality of the fit of the HRG distribution
to data is evaluated in terms of the FVU.
in the 5th column, the difference between $\theta_{CP}$ and $\theta_{HRG}$ is characterized by the ratio of $\Delta=\theta_{HRG}-\theta_{CP}$ to its standard deviation in the 6th column. } 
 \begin{tabular}{|c|c|c|c|c|c|c|} \hline
$\sqrt{s_{NN}}$, GeV  & $\theta^a_{STAR}$  & $\theta^{\rm{old}}_{CP}$ & $\theta_{CP}$  & $\theta_{HRG}$  & FVU$/10^{-4}$ & $|\langle \Delta\rangle| /\sigma_\Delta$ \\ 
       &       & &    &     &  & \\ \hline
7.7    & 2.758 & ---            & ---        & 1.673(56)  & 0.66 & --- \\ 
11.5   & 1.921 &~2.0126752(5)~& ---        & 1.909(35)  & 1.52 & --- \\  
14.5   & 1.741 & ---          & ---        & 1.551(17)  & 1.09 & --- \\ 
19.6   & 1.222 & 1.1643(47)   & 1.142(33)  & 1.2559(68) & 2.04 & 3.4 \\ 
27.0   & 0.929 & 0.8918(21)   & 0.9057(28) & 0.9364(44) & 0.99 & 5.9 \\ 
39.0   & 0.659 & 0.63306(63)  & 0.6588(10) & 0.6623(12) & 0.71 & 2.2 \\ 
54.4   & 0.519 & ---          & 0.5018(9)  & 0.5042(8)  & 0.46 & 2.0 \\ 
62.4   & 0.437 & 0.42773(18)  & 0.4415(13) & 0.4360(16) & 0.78 & 2.7 \\ 
200.0  & 0.170 & 0.161260(74) & 0.1687(3)  & 0.1662(28) & 0.19 & 0.9 \\ \hline
 \end{tabular}
 \label{tab:2}
\end{table}

It should be emphasized  that the Fraction of Variance Unexplained (FVU) in the case of fit function (\ref{eq:5}) is by $2\div 3$ orders of magnitude smaller than FVU for the fit by Gaussian or by the probability density function 
of the free fermion gas.

There is an excellent agreement between $\theta_{CP}$, $\theta^a_{STAR}$ and $\theta_{HRG}$ at $\sqrt{s_{NN}}>19$~GeV, where
determination of $\theta_{CP}$ is possible;
$\theta^a_{STAR}$ are the values of $\theta_{STAR}$ presented in \cite{STAR:2021iop}
for all values of collision energy under consideration.
Significant difference between $\theta_{HRG}$ and 
$\theta^a_{STAR}$ at $\sqrt{s_{NN}}=7.7$~GeV
may be considered in the context of 
the study~\cite{Gazdzicki:2026ubm} where
the onset of deconfinement in the fireball 
is assumed as the collision energy 
increases from 8 to 12~GeV.

\begin{figure}[!ht]
\includegraphics[width=0.6\linewidth]{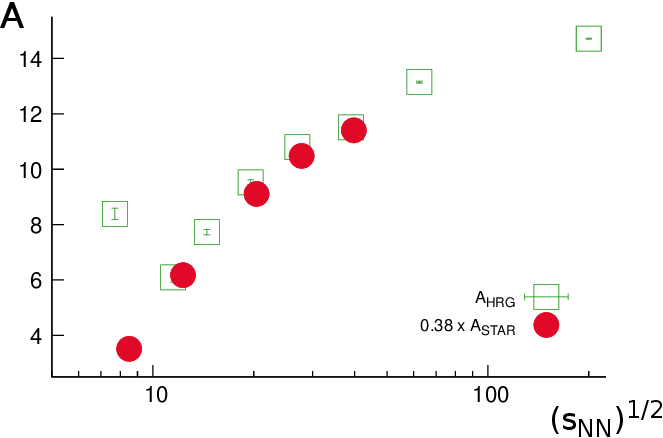}
\caption{The dependence of parameter $A$
  on the collision energy.}
\label{fig:ImgC}
\end{figure}

The dependence of the parameter~$A$ 
determined by the fit formula (\ref{eq:5}),
$A_{HRG}$, on $\sqrt{s_{NN}}$ is shown 
in Fig.\ref{fig:ImgC}. 

An immediate comparison of $A_{HRG}$
with $A_{STAR}=2\sqrt{\langle p \rangle \langle \bar p \rangle}$ is impossible because the acceptance 
in which the data for determination of $A_{HRG}$
were obtained~\cite{STAR:2021iop} differs from that for determination
of $\langle p \rangle$ and $\langle \bar p \rangle$
\cite{STAR:2017sal}.
However, it should be noticed that they coincide up to the constant factor at 
$\sqrt{s_{NN}}>10$~GeV ($A_{HRG}=0.38 A_{STAR}$) 
giving evidence for the validity 
of the HRG-based fit formula~(\ref{eq:5})
at the specified energies.
Since the baryon-density fluctuations 
at $\mu_B=0$ and $140<T<160$~MeV described by the parameter $A$ 
in the HRG model agree well with those evaluated 
in lattice QCD \cite{HotQCD:2012fhj,Vovchenko:2017xad,Bornyakov:2016wld}, the above reasoning gives some evidence for agreement between lattice and experimental data.

\section{Conclusions}

We have first applied the method 
for determining~$\theta$ suggested 
in~\cite{Nakamura:2013ska} 
to determine~$\theta$ and $\mu_B$
from the  STAR data on net-proton 
number fluctuations in a model-independent way.
Our values of $\mu_B$ agree within the error 
limits with those obtained by the STAR 
collaboration on the basis the statistical 
thermal model, while our error in determining 
$\mu_B$ is significantly smaller.

We have proposed a criterion for thermodynamic 
equilibrium of fireballs: the difference of 
$\theta_P$ from a constant function of $P$
is incompatible with thermodynamic equilibrium.

We also have found that fit function based on the HRG
model works well at $\sqrt{s_{NN}}>10$~GeV; no significant deviations from the predictions of this model and, therefore, of lattice QCD have been found in agreement with conclusions of~\cite{Braun-Munzinger:2020jbk}.
Therewith, substantial deviation from the HRG 
model at $\sqrt{s_{NN}}=7.7$~GeV encourages closer examination of the assumptions made here, in particular, the validity of the relation (\ref{eq:5}) at $\sqrt{s_{NN}}=7.7$~GeV as well as the assumption about the production of confined matter in collisions at $\sqrt{s_{NN}}<8 $~GeV made in~\cite{Gazdzicki:2026ubm}.

\vspace{2ex}
\begin{acknowledgments}
We are grateful to V.G.Bornyakov for useful discussions. This research was funded by the Russian Science Foundation (Grant 23 12-00072-P).  
\end{acknowledgments}

\bibliography{bibliography}

\end{document}